\documentclass[twocolumn,dvipsnames,nofootinbib,floatfix,groupedaddress,superscriptaddress]{revtex4-1}
\PassOptionsToPackage{unicode}{hyperref}
\PassOptionsToPackage{bookmarks=false}{hyperref}
\usepackage{braket}
\usepackage[utf8]{inputenc}
\usepackage{bbold}
\usepackage{amsmath}
\usepackage{geometry}
\usepackage{fullpage}
\usepackage{graphicx}
\usepackage{mathrsfs}
\usepackage[newitem,newenum]{paralist}
\usepackage{units}
\usepackage{slashed}
\usepackage{float}
\usepackage{color,xcolor}
\usepackage{bm,paralist,xspace,url}
\usepackage{multirow}
\usepackage{fontawesome}

\graphicspath{{./PlotsPRD/}}

\newcommand{\br}{\mathop{\mathrm{br}}\nolimits}

\newcommand{\cdf}{\mathop{\mathrm{cdf}}}

\usepackage[margin=false,inline=true,index=true,draft]{fixme}
\fxsetup{
  inlineface=\footnotesize
}
\fxusetheme{colorsig}

\FXRegisterAuthor{oleg}{anoleg}{O.R.}
\FXRegisterAuthor{alex}{analex}{A.M.}
\FXRegisterAuthor{kirill}{ankirill}{K.B.}

\usepackage{xspace}

\begin{document}

\title{Unveiling new physics with discoveries at Intensity Frontier}

\author{Oleksii~Mikulenko}
\email{mikulenko@lorentz.leidenuniv.nl}
\affiliation{Instituut-Lorentz, Leiden University, Niels Bohrweg 2, 2333 CA Leiden, The Netherlands}
\author{Kyrylo~Bondarenko}
\affiliation{Nordita, KTH Royal Institute of Technology and Stockholm University, Hannes Alfv\'ens v\"ag 12, 10691
Stockholm, Sweden}
\author{Alexey~Boyarsky}
\affiliation{Instituut-Lorentz, Leiden University, Niels Bohrweg 2, 2333 CA Leiden, The Netherlands}
\author{Oleg~Ruchayskiy}
\affiliation{Niels Bohr Institute, University of Copenhagen, Blegdamsvej 17, DK-2010, Copenhagen, Denmark}

\date{8 December 2023}

\begin{abstract}
  The idea of feebly interacting particles (FIPs) has emerged as an important approach to address challenges beyond the Standard Model. The next generation of Intensity Frontier experiments is set to explore these particles in greater depth. While many experiments may detect FIP signals in unexplored regions of masses and couplings, interpretation of the properties of particles behind these signals is typically neglected. In this paper, we present a novel framework designed to systematically determine the models behind a potential signal. Our approach allows us to assess the scientific reach of experiments beyond the concept of sensitivity to the smallest coupling constant leading to a detectable signal. We clarify the potential impact such signals could have on particle physics models. This paper is complemented by a Python package for the presented framework, available at \faGithub\href{https://github.com/omikulen/modeltesting}{ omikulen/modeltesting}.
   
\end{abstract}
\maketitle

\section{Introduction}
\label{section:intro}
The Standard Model of particle physics provides a thorough understanding of fundamental particles and their interactions. Its validity and precision have been affirmed through continuous operations and meticulous data analyses at major experimental facilities worldwide. Despite its significant achievements, certain phenomena and anomalies remain outside of its scope. These \emph{beyond the Standard Model} (BSM) problems include the true nature of dark matter, the origin of neutrino masses, and the underlying causes of baryon asymmetry in the Universe.  
These puzzles suggest new sectors of physics that are yet to be uncovered.
The absence of signals from new particles at the electroweak scale has pivoted the community's attention toward alternative avenues for BSM physics. Over the past decade, the concept of ``feebly interacting particles'' (FIPs) has garnered increasing interest as a potential paradigm for addressing these challenges~\cite{Antel:2023hkf, Beacham:2019nyx}. This perspective includes hypothetical particles with masses at the GeV-scale with the interaction strength more feeble than the weak force. 
These properties make FIPs a prime candidate for exploration using modern experimental techniques and significantly fueled the advancement of the Intensity Frontier program.

The list of the new proposed Intensity Frontier experiments includes beam dump facilities, such as SHiP~\cite{Alekhin:2015byh,SHiP:2015vad,Aberle:2839677}, SHADOWS~\cite{Baldini:2021hfw,SHADOWS_BDF}, and HIKE$_{\text{dump}}$~\cite{CortinaGil:2839661} at CERN or DUNE~\cite{DUNE:2020ypp,DUNE:2020fgq} and DarkQuest~\cite{Batell:2020vqn} at Fermilab;
or experiments alongside LHC, featuring MATHUSLA~\cite{MATHUSLA:2019qpy} and FACET~\cite{Cerci:2021nlb} at the CMS site; FASER~\cite{FASER:2018bac}, SND@LHC~\cite{SHiP:2020sos,SNDLHC:2022ihg},  AdvSND, FASER2, and ANUBIS~\cite{Bauer:2019vqk} associated with ATLAS; CODEX-b~\cite{Aielli:2019ivi} accompanying LHCb and AL3X~\cite{Dercks:2018wum} at ALICE.

This proliferation of the Intensity Frontier experiments has been aimed at maximizing the sensitivity of detecting a signal across as wide a range of models as possible. Typically, the science reach of Intensity Frontier projects has been assessed by their capacity to advance deeper into an unexplored parameter space.

In this paper, we contend, however, that the genuine ``discovery potential'' should be evaluated by the insights that can be derived should a signal be \emph{detected}. 
Many of the Intensity Frontier experiments have the potential to observe not \emph{few} but hundreds or even thousands of events over the lifetime of the experiment.
This plethora of data opens up the possibility of studying the particle ``behind the signal'' in depth. 
One can determine not only the basic particle's properties like mass and spin but eventually progress towards evaluating whether this newfound particle might hold the key to unresolved BSM phenomena.

\emph{Our paper is devoted to the systematic exploration of this question and expands the companion paper}~\cite{PRL}.
We present a systematic framework that allows \textit{(1)} to confront any discovered signal with the wide variety of BSM models and to assess their compatibility; and \textit{(2)} to evaluate the discovery potential of Intensity Frontier experiments based on the number of events they can accumulate and the types of signal they can detect.  The paper discusses both the approach and the details of the implementation. Regarding the second question, we present an analysis of the potential discovery of a sterile neutrino, commonly referred to as heavy neutral leptons (HNL).
These particles emerge as \emph{neutrino portal} -- renormalizable interaction able to explain neutrino masses and oscillation (a famous Type-I seesaw model \cite{Minkowski:1977sc,Yanagida:1979as,Glashow:1979nm,Gell-Mann:1979vob,Mohapatra:1979ia,Mohapatra:1980yp,Schechter:1980gr,Schechter:1981cv}).
Two such particles with masses in the GeV range can also provide the mechanism of generation of the baryon asymmetry of the Universe~\cite{Akhmedov:1998qx,Ghiglieri:2020ulj,Klaric:2020phc,Klaric:2021cpi,Asaka:2005pn}.
In the framework of the \textit{Neutrino Minimal Standard Model}~\cite{Asaka:2005an,Asaka:2005pn,Boyarsky:2009ix} they can also create initial conditions for the generation of dark matter in terms of the lighter (keV-scale) sterile neutrino in the early Universe~~\cite{Boyarsky:2018tvu, Eijima:2020shs}.
The two heavy HNLs must be mass degenerate to account for the baryon asymmetry and to explain neutrino masses and oscillations. These conditions place constraints on the interactions of the HNL particle with different lepton flavors. Such constraints offer a valuable avenue for examination, for they allow us to assess whether the observed signal could indeed be attributed to a pair of degenerate HNLs, responsible for neutrino oscillations. For high-energy facilities, this question was studied in~\cite{Drewes:2017zyw}.

The structure of the paper is as follows. In Sec. 2, we focus on the general identification of particles, elaborating on how to differentiate between commonly considered portal models. In Sec. 3, the first part delves into the phenomenology of HNLs and provides the rationale behind the method chosen for their study; the second part outlines a more general method that has broader applicability. Finally, we present our findings in Section 4 and offer conclusions in Section 5.

\section{Determining the nature of the signal}

Whenever a potential signal has been detected, the questions arise: 
\begin{compactenum}[(a)]
    \item What is the mass and the spin of the particle?
    \item What is the strength of its interaction with the SM particles?
    \item Is the observed particle consistent with a given FIP model?
    \item How does it relate to the BSM phenomena?
\end{compactenum}
The answer to these questions is a multi-stage process. 
Below, we show step by step how each of these questions can be answered in the case of the three main FIP models, described by renormalizable interactions (or \textit{portals}) with the SM sector \cite{Alekhin:2015byh,Beacham:2019nyx,Agrawal:2021dbo}, as well as the most famous nonrenormalizable interaction, an axion-like particle. 

In this section, we discuss the determination of the mass and spin of new particles.

Accurate mass measurement requires the particle to decay into a fully detectable state. 
The decays into reconstructable final states -- light leptons ($e^\pm$ and $\mu^\pm$) and long-lived mesons $\pi^{\pm}$, $K^\pm$ allows to recover the mass as the invariant mass of such final states. 
An important aspect here is the ability of the experiments to identify photons and to measure their kinematics. Such detector systems (usually electromagnetic calorimeters, ECALs) play a vital role in reconstruction decays into neutral mesons, e.g. $\pi^0 \to 2\gamma$, as well as identification of axion-like particles, see e.g.\ \cite{SHiP:2015vad,SHiP_CDR}.

A direct measurement of the spin of a new particle is not feasible, as it requires a measurement of the polarizations of the final state products. Therefore, if the decaying particle is a \textit{boson}, one can deduce it indirectly, in a \textit{model-dependent way}, as discussed below. 
For a new \textit{fermion}, however, this is not an obstacle: the final state is required to have an odd number of fermions, contrary to the case of the decay of a boson. An example is a fully detectable decay into $X \to l \pi^+$, which is a smoking gun signature for establishing the fermionic nature of the new particle (given that the absence of any missing particle is firmly established). Other options include a decay with missing energy $X \to ll\nu$, assuming that the invisible particle is a neutrino. These decays, however, are more ambiguous, for they can be naturally present in larger extensions of the SM with several feebly interacting particles $X$, $X'$, in the form of the decay $X\to l l X'$. 

Having discussed general scenarios, we focus on distinguishing various commonly considered \textit{portal} models~\cite{Beacham:2019nyx}, incorporating a scalar, fermion (heavy neutral lepton, HNL), vector, and pseudoscalar (axion-like, ALP) particle. We first note that in the target region of $\sim\unit[1]{GeV}$ mass, most of these models predict  $O(1)$ branching ratios into fully reconstructable states, with the exceptions of purely $\gamma$-coupled ALP and $\tau$-coupled HNL. Therefore, we conclude that up to ten of signal events would suffice for mass measurement. Hence, in what follows, we assume that the mass of the particle is known.

The coupling constants in a given model are, of course, connected to the total number of events we observe. The relationship gets complicated when the model has many couplings. In Section~\ref{sec:HNL}, we will go into more details in the HNL case.
For the rest of this section, we will take a quick look at the models that have only one coupling constant. In these instances, the total number of events, along with the particle's mass, is uniquely related to the coupling constant at a given experiment. 
For every model, it is important to distinguish between the main channels, which are used to measure the interaction strength, and the subdominant consistency channels, which help to confirm that the observed particle fits with the model in question.

\begin{table}[t!]
    \centering
    \begin{tabular}{|l|c|}
    \hline
         \textbf{Particle} & \textbf{Decay modes}  \\
    \hline
         \multirow{3}{*}{HNL} & $l_\alpha h^\pm$ \\
         &  $l_\alpha l_\beta  +\text{inv.}$
         \\
         & $h + \text{inv.}$  \\
    \hline
         Scalar & $l_\alpha \bar l_\alpha$,\quad  $h$ \\
    \hline
         Vector & $l_\alpha \bar l_\alpha$,\quad $h$ \\
    \hline    
         ALP (photon coupling) & $\gamma\gamma$ \\
    \hline  
         ALP (gluon coupling) &  $h$ \\
    \hline  
         ALP (fermion coupling) & $l_\alpha l_\alpha$,\quad $h$ \\
    \hline
    \end{tabular}
    \caption{The expected final states for each of the portal models. The notation $h$, $h^+$ represents hadronic states with total zero and $\pm 1$ electric charge, respectively. These hadronic states include $\pi^{\pm}$, $K^{\pm}$, and neutral mesons, which decay instantaneously, predominantly into photons.}    \label{tab:decay_modes_allmodels}
\end{table}

In our discussion below, we focus on the mass region of the new particle from $\unit[0.2]{GeV}$ to $\unit[2]{GeV}$. At smaller masses, most of the interesting decays are kinematically forbidden. At larger masses, multihadron and chain decays are possible, necessitating a careful analysis of the reconstruction capabilities of a given detector system.

\begin{asparadesc}
    \item[Scalar $S$] mixes with the Higgs boson, implying a very distinct hierarchy of interactions with different fermions. As a result, it decays exclusively into the most massive kinematically accessible pair ($ee$, $\mu\mu$, $\pi\pi$, $KK$, \dots) at scalar masses of interest~\cite{Boiarska:2019jym}. 
    In hadronic decays, the isospin symmetry leads to the ratio $2:1$ for decays into $\pi^+\pi^-:2\pi^0$ and $1:1$ for $K^+K^-:K^0\bar K^0$.

    \item[HNL $N$] mixes with neutrino, violates the \textit{visible} conservation of lepton number. Decays in $e\mu$, $l^{\pm}+\text{hadrons}$ are unique to the neutrino portal and dominant around the $\sim\unit[1]{GeV}$ mass~\cite{Bondarenko:2018ptm} (except for the pure $\tau$ mixing scenario).
    \item[Vector $A'$] couples to the charge of the SM particles. Above the muon decay threshold $m_{A'} > 2m_\mu$, the branching ratios to electron and muon pairs are equal and not suppressed $\text{Br}\gtrsim 10\%$~\cite{Ilten:2018crw}. In hadronic decays, the $2\pi^0$ final state is forbidden by conservation of angular momentum, which serves as a distinctive feature of a vector-particle decay.
    \item[ALP ($\gamma$-coupled)] decays solely into a photon pair. 
    \item[ALP ($g$-coupled)] mixes with neutral $\pi^0$, $\eta$ and $\eta'$-mesons, gaining the same decay patterns~\cite{Aloni:2018vki}. No leptonic final states $ee$, $\mu\mu$. Remarkably, the decays into $2\pi$, $4\pi$, and $2K$, exemplary for scalar and vector particles, are absent because of the conservation of CP.
    \item[ALP ($f$-coupled)] in the scenario of equal coupling to leptons and quarks, the decays are dominated by $2\mu$ final states in the mass range 0.2 --- 1 GeV, which is a unique signature of this model~\cite{Garcia:2023wzp}. At larger masses up to $\unit[2]{GeV}$, the decay patterns resemble that of a gluon-coupled ALP. However, muonic decays remain not too suppressed $\text{Br}\gtrsim 1\%$ to serve as a distinctive feature. 
\end{asparadesc}

To conclude, an experiment equipped with detectors of $e$, $\mu$, $\pi^\pm$, $K^\pm$, along with photonic detectors, can differentiate between the majority of the models discussed here with as few as tens of events. This can be achieved by focusing on decay modes that are suppressed or forbidden in other models.

\section{Example: probing the seesaw mechanism}
\label{sec:HNL}

After revealing the fundamental properties of a new particle --- its mass and spin, the more subtle details of the underlying model can be addressed. To begin with, it is clear that the number of observed events is linked to the overall coupling of the new physics sector to the SM. To list a few more nuanced questions: a) to what extent a new particle is consistent with the assumption of the portal model; b) what are the mixing angles of an HNL to each leptonic flavor; c) how to study the generic axion model, with all the listed couplings treated on the equal footing. 

In our analysis, we used the set of possible final states of the decay of the examined particles. At this step, it is necessary to identify the relevant \emph{classes} of the decay modes that contain the relations between the model parameters of interest. Within a class, the branching ratios are fixed by the model structure, and their analysis may only shed light on the overall \textit{ consistency} of the model with the data.

In the following section, we focus on the analysis of the model of two HNLs, which may be related to neutrino oscillations through the seesaw mechanism. This section consists of two parts. In the first part, we discuss details that are relevant to probing HNLs specifically and the motivation behind the method employed in the paper. In the second part, we present the general method, which can be used for a variety of similar tasks to differentiate models of new physics particles based on the theoretical branching ratios of their decays.

\subsection{HNL-related details}

As mentioned in the introduction, we want to study the potential of probing the specific model of two mass-degenerate HNLs that can be responsible simultaneously for neutrino masses and the generation of baryon asymmetry. The two species provide two active neutrinos with masses, leaving the lightest one massless.

We will consider the region of HNL masses between $0.5\lesssim m_N \lesssim  2$ GeV. The reason is the following: on the one hand, the parameter space not excluded below the kaon mass $m_N\lesssim 0.5$ GeV allows little room for a new particle~\cite{Bondarenko:2021cpc}. On the other hand, we avoid the additional complexity in the interpretation of the observed decays, which appears for $m_N \gtrsim 2$ GeV. This includes reconstructing decays of $\tau$-lepton and estimating the branching ratios of decays into observable final states through heavy mesons $N\to D$. A comprehensive review of the phenomenology of GeV-scale HNLs can be found in~\cite{Bondarenko:2018ptm}.

Heavy neutral leptons interact with the SM model through mixing with active neutrinos. For each lepton flavor $\alpha$ and HNL species, there is a separate mixing angle $U_{\alpha I}$ that determines the strength of the corresponding interaction. Given the double-edged nature of such mixing, which is the origin of active neutrino oscillations in the first place, it is possible to relate the quantities $U_{\alpha I}$ to the neutrino oscillation data (the PMNS matrix and neutrino masses) using the Casas-Ibarra parametrization~\cite{Casas:2001sr}. 

The lower limit on the overall interaction strength, represented by the total coupling constant $U^2 = \sum_{\alpha I} U^2_{\alpha I}$, is given by the seesaw bound:
\begin{equation}
    U^2_\text{seesaw} = \frac{\sqrt{\Delta m^2_\text{atm}}}{m_N} = 5 \cdot 10^{-12}\frac{\unit[1]{GeV}}{m_N}.
\end{equation}
In the region that can be probed by future experiments~\cite{Abdullahi:2022jlv}, the total coupling constant $U^2$ exceeds the seesaw limit. To prevent active neutrino to acquire too large masses, the two mass degenerate HNLs, if detected, must exhibit an \textit{approximate lepton symmetry}, and hence have the same mixing angles $U^2_{\alpha 1} \approx U^2_{\alpha 2} \approx  \frac{1}{2} U^2_{\alpha}$. 

In the approximate lepton symmetry regime, the scale of the mixing angle $U^2 = \sum_\alpha U^2_\alpha$ is not bounded from above and can be arbitrary, but the relative contributions of different lepton flavors, the \textit{mixing ratios}:
\begin{equation}
    x_\alpha = U^2_\alpha /U^2, \qquad \sum_\alpha x_\alpha = 1,
\end{equation}
can be constrained~\cite{drewes:2018gkc, Chrzaszcz:2019inj, Bondarenko:2021cpc}, as shown in Fig.~\ref{fig:neutrino_priors}. For two massive active neutrinos, the PMNS matrix has a single Majorana phase $\eta$, whose variation creates an ellipse in the ternary plot. The uncertainties in the other neutrino oscillation parameters deform the ellipse, producing the shown regions. For the details, see Appendix~\ref{sec:neutrino_priors}.

\begin{figure}[h!]
    \centering
    \includegraphics[width = \linewidth]{PlotsPRD/nu_priors.jpg}
    \caption{Prior probability distributions in the $x_\alpha = U^2_\alpha/U^2$ plane for the normal (red) and inverted (blue) hierarchies. The levels correspond to 75\%, 50\%, 25\%, and 10\% from the darkest to the brightest. The stars represent the benchmark models from~\cite{Drewes:2022akb}. The probabilities were computed with the NuFit 5.2 data, see text and Appendix~\ref{sec:neutrino_priors} for details. }
    \label{fig:neutrino_priors}
\end{figure}

The branching ratio of decay into some final state $N\to X$ depends only on the mixing ratios $x_\alpha$ but not on the overall scale $U^2$:
\begin{multline}
\label{eq:BrDef}
    \text{Br}(N\to X) = 
    \\
    = \frac{x_e \Gamma_e(X) + x_\mu \Gamma_\mu(X) + x_\tau \Gamma_\tau(X)}{x_e \Gamma_e + x_\mu \Gamma_\mu + x_\tau \Gamma_\tau}
\end{multline}
where $\Gamma_\alpha = \Gamma/U_\alpha^2$ is the decay width for an HNL that mixes purely with the $\alpha$ neutrino flavor, normalized to a unit mixing angle.

\subsection{Outline of the general method}

To perform an analysis of the sensitivity, we have built a general framework to differentiate between models that are parametrized by branching ratios. Technical details are given in Appendix~\ref{sec:technical_appendix}.

For each model $\mathcal M$ of new physics particles, we define two quantities:
\begin{enumerate}
    \item the expected total number of events, $N$, happening within the acceptance of an experiment (e.g., its decay volume)
    \item the set of $n$ observable decay channels, numerated by an index $i = \overline{1, n}$, with corresponding branching ratios $\mathrm{Br}_i$.
\end{enumerate}

The average number of events in each observable channel is given by
\begin{equation}
    \lambda_i = N \times \br_i + b_i
\end{equation}
where $b_i$ is the expected background, while the \emph{effective branching ratio} $\br_i$ is defined as the product of the branching ratio, $\mathrm{Br}_i$, and the mean detection efficiency for that specific channel,  $\epsilon_{\text{det}, i}$.
The details of a specific experiment are encoded into the quantities $\epsilon_{\text{det}, i}$ and $b_i$.

For a specific dataset of the observed counts in each channel $s_i$, we define the rejection criteria using the $\chi^2$-test. The model is accepted at the confidence level CL if
\begin{equation}
    \label{eq:rejection_criteria}
     1- \text{CL} > \mathscr{P}(\mathcal M) \times \bigg[1-\cdf(\chi^2_\lambda = \chi^2(s_i|\lambda_i) )\bigg]  
\end{equation}
where $\cdf(\chi^2_\lambda)$ is the cumulative distribution function for a random variable $\chi^2_\lambda \equiv \chi^2(s^{\lambda}_i|\lambda_i)$, with the data $s^\lambda_i$ drawn from the Poisson distribution with the mean $\lambda_i$. The quantity $\mathscr P(\mathcal M)$ is the prior probability of the model, which incorporates constraints from independent experiments. Note that the smaller $\mathscr P$ is, the less tension in the experimental results $\cdf(\chi_\lambda)$ we need to observe to reject the model.

For a model $\mathcal M_r$, assumed to be the real model of nature, the experimental signal is a random realization of the expected theoretical predictions. Due to this randomness, we cannot predict how much the actual signal will be in tension with the tested model. Therefore, it is reasonable to measure the sensitivity of discrimination between the models as the \textit{exclusion probability} $\mathcal P_\text{exc}(\mathcal M_t | \mathcal M_r)$: the probability that the experimental results originating from $\mathcal M_r$ would exclude the tested model $\mathcal M_t$, using the rejection criteria~\eqref{eq:rejection_criteria}.

We are interested in the discrimination between two physical models that predict different branching ratios $\text{Br}$. The total number of events in the tested model $N_t$ is a free parameter that must be fitted by the data. Therefore, in the rejection criteria, the best-fit value of $N_t$ should be used that minimizes $\chi^2(s_i|\lambda_i)$.

By increasing the number of events $N_r$ in the real model, we can measure the branching ratios with better precision. This means that the \textit{exclusion probability} increases with $N_r$. At this step, we can finally define the \textit{sensitivity to differentiate between the real model $\text{Br}_r$ and a tested model $\text{Br}_t$}: It is the minimal number of $N_r$ that allows discrimination between the two models at CL with the exclusion probability, exceeding some predefined threshold $P$. In our analysis, we adopt the values CL = 90\%  and $P = 90\%$.

\section{Results}

In this work, we consider mixing ratios $(x_e, x_\mu)$ that are inconsistent with normal/inverse hierarchies and estimate the number of events required to exclude any model consistent with NH/IH. 

For our analysis, the relevant decay modes are given in Tab.~\ref{tab:decaymodes}.  Except for $N\to 3\nu$, all decay modes are potentially detectable and distinguishable. However, depending on the capabilities of the detection systems, some final states may be missed. As mentioned earlier, the status of the photonic detectors is not clear. This may be crucial for observing the decay $N \to \nu \pi^0 \to \nu (2\gamma)$ that dominates the decay channel 4.

\begin{table*}[t!]
    \centering
    \begin{tabular}{|c|c|c|c|}
    \hline
        &decay mode& mixing & $\Gamma_\alpha \times 10^{13}$, GeV   \\
    \hline
        0)&$N\to 3\nu$& $U^2_{e,\mu,\tau}$ &  1.7  \\
    \hline
        1)&$N\to \nu e e$ & $(U^2_e$, $U^2_{\mu,\tau})$ & $(1.0, 0.2)$ \\
    \hline
        2)&$N\to \nu e \mu$   & $U^2_{e,\mu}$& $1.7 $ \\
    \hline
        3)&$N\to \nu \mu \mu$   & $(U^2_\mu$, $U^2_{e,\tau})$ & $(1.0, 0.2) $ \\
    \hline
        4)&$N\to \nu h^0$ (NC)& $U^2_{e,\mu,\tau}$ & 2.5 \\
    \hline
        5)&$N\to e h^+$  (CC) & $U^2_e$ & 5.0 \\
    \hline
        6)&$N\to \mu h^+$  (CC) & $U^2_\mu$ & 5.0 \\

    \hline 
    \end{tabular}
    \caption{The relevant decay modes for establishing the ratios of mixing angles $x_\alpha = U_\alpha^2/U^2$. The column ``mixing'' represents the neutrino flavors that mediate the decay. The notation $U^2_{\alpha, \beta\dots}$ highlights the equal contribution of mixing with $\alpha,\beta\dots$ to the decay width. For the specific mass $m_N = \unit[1.5]{GeV}$, the decay widths entering~\eqref{eq:BrDef} are given to illustrate the relative importance of different channels.}
    \label{tab:decaymodes}
\end{table*}

\begin{figure}[h!]
    \centering
    \includegraphics[width=\linewidth]{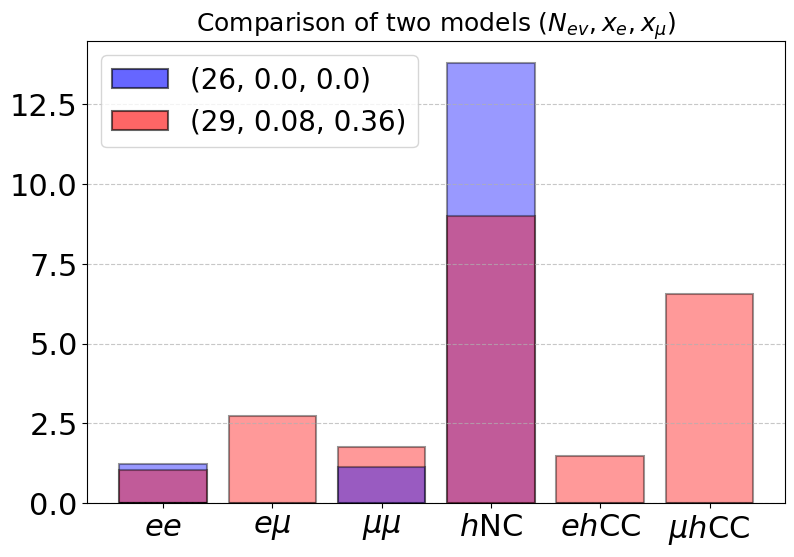}
    \caption{Predictions $\lambda_i$  for the different channels ($m_N = 1.5$ GeV).
   The true model with $\tau$-only mixing is shown in blue. The absence of the signal in $\ell h$ channels allows us to determine that $x_\ell = 0$, while $e\mu$ channel verifies this. At the same time, the overall scale of $U^2_\tau\sim U^2$ is determined from the \textsc{nc} channel and confirmed by $ee$ and $\mu\mu$ channels.  The best-fit model consistent with NH is shown in red.}
    \label{fig:dominant_channel}
\end{figure}

Different decay modes provide different information, see Fig.~\ref{fig:dominant_channel} for illustration.  The hadronic decays
mediated by the charged current (CC) $N\to eh^+$, $\mu h^+$ probe directly mixing with the electron and muon neutrinos and rely only on the identification of the associated charged lepton and observation of the final hadrons. The decays into hadrons mediated by the neutral current (NC) $N\to \nu h^0$ are insensitive to mixing couplings because the flavor of the neutrino is not observed. These decays provide information on the total normalization of the branching ratios and thus indirectly probe $U^2_\tau$. The decay into $N\to l_\alpha\bar l_\alpha$ can be mediated by any mixing. However, decays mediated by the same flavor $U_\alpha$ mixing can proceed through the CC interaction, which yields a larger decay width by a factor $\frac{1+4\sin \theta^2_W + 8\sin^4 \theta_W}{1-4\sin \theta_W^2 + 8\sin^4 \theta_W}\simeq 4.7$ compared to the NC decays, assuming the mass of HNL is far from the kinematic threshold $m_N> 2 m_l$. Finally, decays $N\to \nu e \mu$ occur by mixing with $e$, $\mu$ with the same decay width dependence.

\begin{figure}[h!]
    \centering
    \includegraphics[width = 0.8\linewidth]{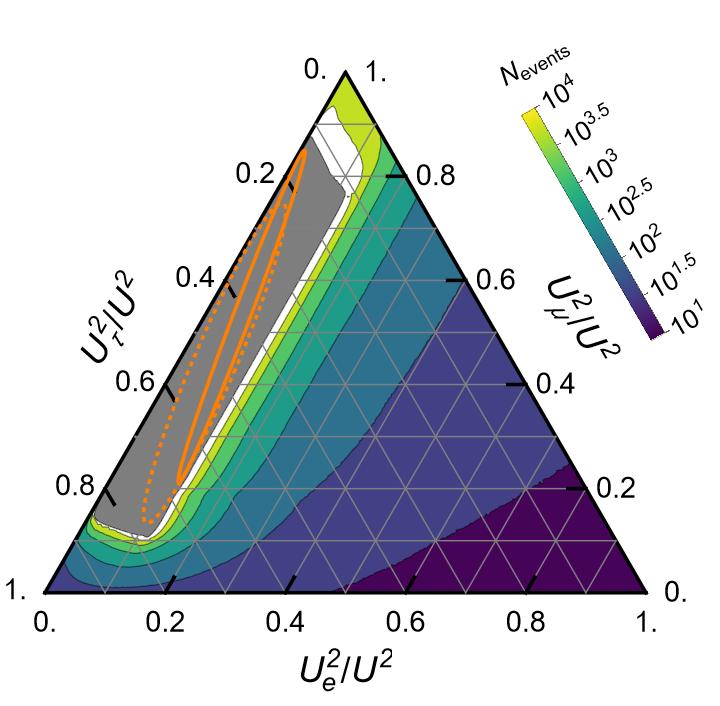}
    
    \includegraphics[width = 0.8\linewidth]{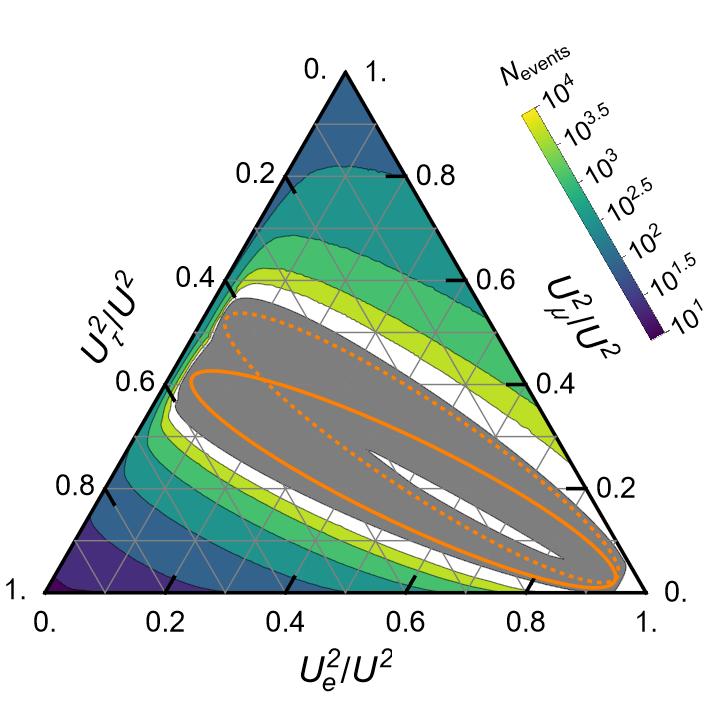}
    \caption{The number of events required to rule out the normal (left) or inverted (right) hierarchy, for a given point $(x_\alpha)$. HNL mass considered is $m_N =\unit[1.5]{GeV}$}
    \label{fig:hierarchy_exclusion_maps}
\end{figure}

The number of events refers to all events that occur in the decay volume, including invisible decays in $3\nu$. We assume unit detection efficiency and zero background for all visible channels (1-6) of Tab.~\ref{tab:decaymodes}, including the decays (4) with photons in the final state.

The results for an HNL with mass $m_N = \unit[1.5]{GeV}$ are shown in Fig.~\ref{fig:hierarchy_exclusion_maps}. The required number of events to probe the benchmark models of~\cite{Drewes:2022akb} is presented in Tab.~\ref{tab:benchmark_exclusions}. Observation of $10^3$ events would allow us to exclude most of the ternary plot inconsistent with one of the neutrino hierarchies.

\begin{table}[h!]
    \centering
    \begin{tabular}{|c|c|c|c|c|}
         \hline
         $U_e^2:U_\mu^2:U_\tau^2$& NO all & IO all & NO w/o $\nu h^0$& IO w/o $\nu h^0$ \\
         \hline
         1:0:0 & 40 & 25 & 80 & 70 \\
         \hline
         0:1:0 & 2500 & 100 & 4000 & 200\\
         \hline
         0:0:1 & 15 & 20000 & 25 & 100\,000\\
         \hline
         0:1:1 & $\gtrsim$ 5000 & 400 &  5000 & 400 \\
         \hline
         1:1:1 & 140 & --- & 500 & --- \\
         \hline
    \end{tabular}
    \caption{The required numbers of events to exclude NH/IH for several benchmark models, which are shown as stars in Fig.~\ref{fig:neutrino_priors}. The mass of HNL is assumed $m_N = \unit[1.5]{GeV}$.}
    \label{tab:benchmark_exclusions}
\end{table}

If two HNLs constitute the solution to neutrino masses, the signal observation enables a probe of the properties of active neutrinos. In particular, it offers a unique opportunity for the determination of the Majorana phase. A precise measurement of the phase provides a concrete prediction and target for future neutrinoless double beta $0\nu\beta\beta$ experiments~\cite{Drewes:2016jae,Drewes:2016lqo}. Confirming this prediction or disproving it would become a win-win situation, whether giving strong evidence for the model or ultimately rejecting it.

\begin{figure}[t]
    \centering
    \includegraphics[width = \linewidth]{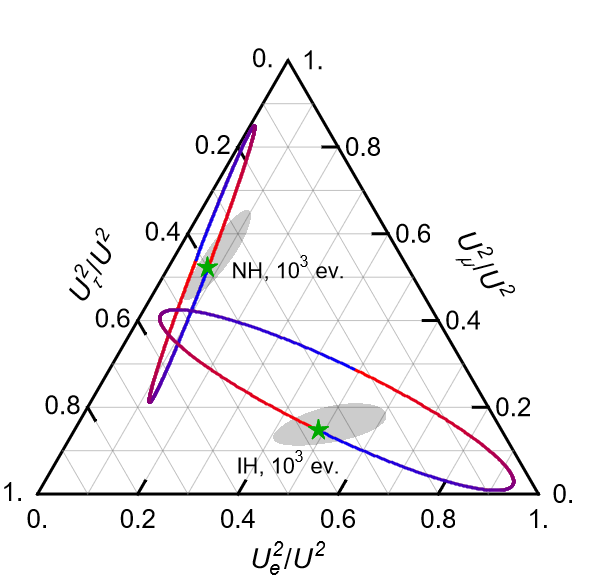}
    \includegraphics[width = \linewidth]{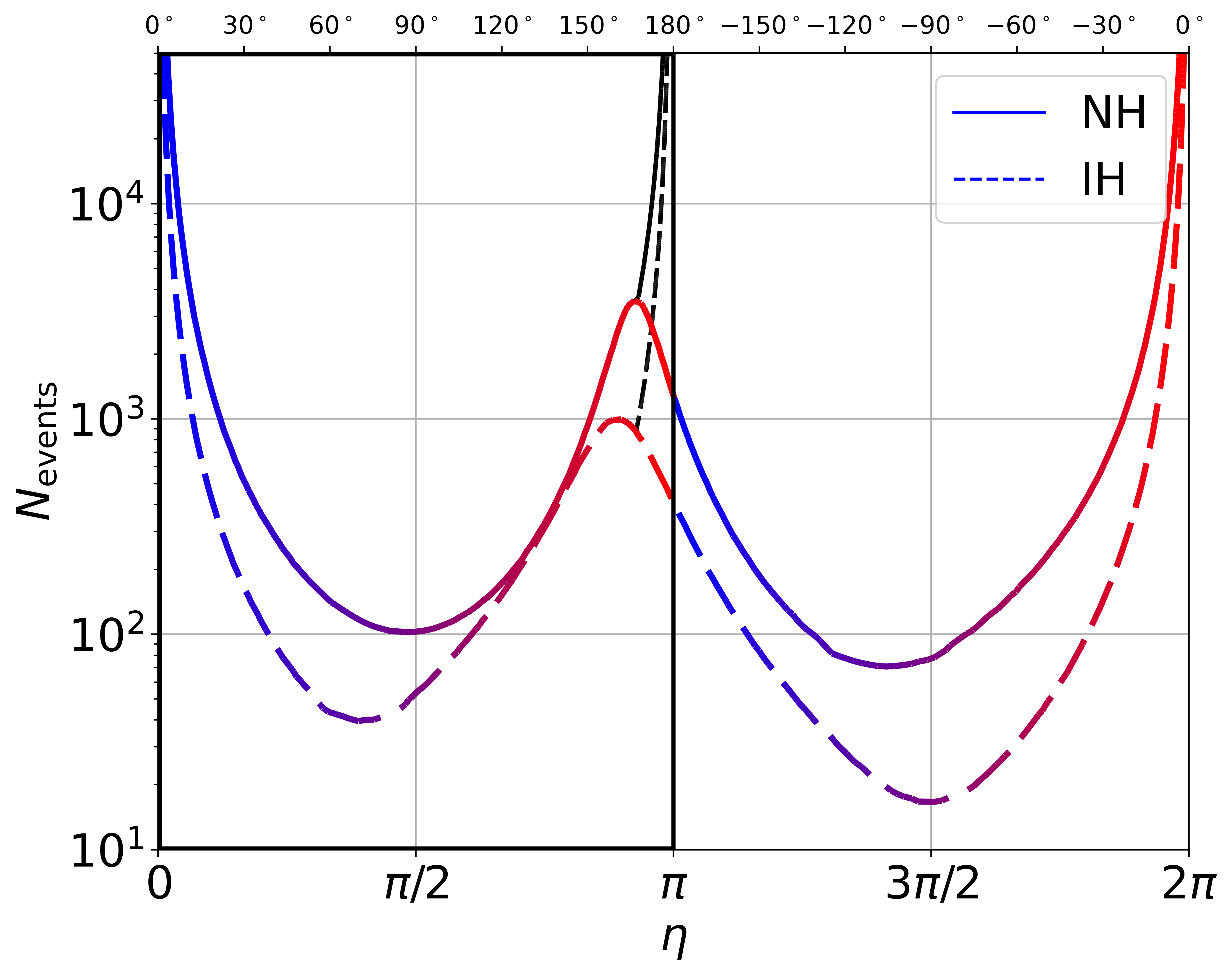}

    \caption{Determining the Majorana phase of active neutrinos for two benchmark models, depicted as stars in the ternary plot and corresponding to $\eta=0$ for NH or IH. The Majorana phase $\eta\in [0,\pi)$ runs from brown ($\eta=0$) to green ($\eta = \pi$).
    The main plot shows the required number of events needed to rule out other points on the ellipse (colored, full range) and Majorana phase (black, in the frame), see text for details. The superior sensitivity to the inverse hierarchy arises due to the more dispersed region of ($x_e$, $x_\mu$) compatible with neutrino oscillations.}
    \label{fig:majophase}
  \end{figure}

Figure~\ref{fig:majophase} demonstrates the potential of measuring the Majorana phase for a benchmark model, represented by a green star ($\eta = 0$), under the assumption that other neutrino parameters are determined with high precision (for example, after 15 years of the DUNE operation, see \cite{DUNE:2020ypp}). 
For each point in the $x_\alpha$ space, two Majorana phases $\eta$ and $\eta + \pi$ are possible, depending on the concrete way of lepton symmetry violation in the HNL sector. Therefore, we define the phase in the range $\eta\in[0,\pi)$. The color shows the change of $\eta$ from 0 (blue) to $\pi$ (red). In the bottom panel, the colored lines show the sensitivity for the determination of the point along the ellipse. The black lines represent the sensitivity to $\eta$ in the range $(0,\pi)$, accounting for the degeneracy of the values. It is possible to determine the phase $\eta$ with a precision of about $50^\circ$ by observing $\mathcal{O}(10^3)$ events.

The number of events can be converted to sensitivity in terms of the coupling constant $U^2$. We use the \textsc{SensCalc} package~\cite{Ovchynnikov:2023cry} to calculate the corresponding values of $U^2$. For sufficiently long-lived particles, namely:
\begin{equation*}
    U^2 \ll \frac{(\unit[100]{m})^{-1}}{C\frac{G_F^2 m_N^5}{192 \pi^3}} \approx 10^{-5}\left(\frac{\unit[1]{GeV}}{m_N}\right)^5
\end{equation*}
where $C$ is a numerical factor of order $1$, the HNL decay length exceeds the typical scales of the experiment. In this scenario, the probability of the HNL decay scales linearly with $U^2$, and the overall dependence of the number of events on the mixing angles takes the form $N_\text{events} \propto (\sum a_\alpha  U^2_\alpha) \times (\sum b_\beta  U^2_\beta)$~\cite{Bondarenko:2019yob}. Given this, one has to only know six independent functions $N_\text{ev}(U^2)$ for different combinations of mixing ratios $x_\alpha$ to build the full function. We use the following parametrization:
\begin{multline*}
    N_\text{events}(m_N, U^2, (x_\alpha)) = \\
    \sum_\beta x_\beta(2x_\beta -1) N_\text{ev.}(m_N, U^2, x_\beta=1) +\\
    \sum_{\beta \neq \delta}  x_\beta x_\delta N_\text{ev.}(m_N, U^2, x_\beta=x_\delta=\frac{1}{2})
\end{multline*}

\begin{figure}[t]
    \centering
    \includegraphics[width = 0.85\linewidth]{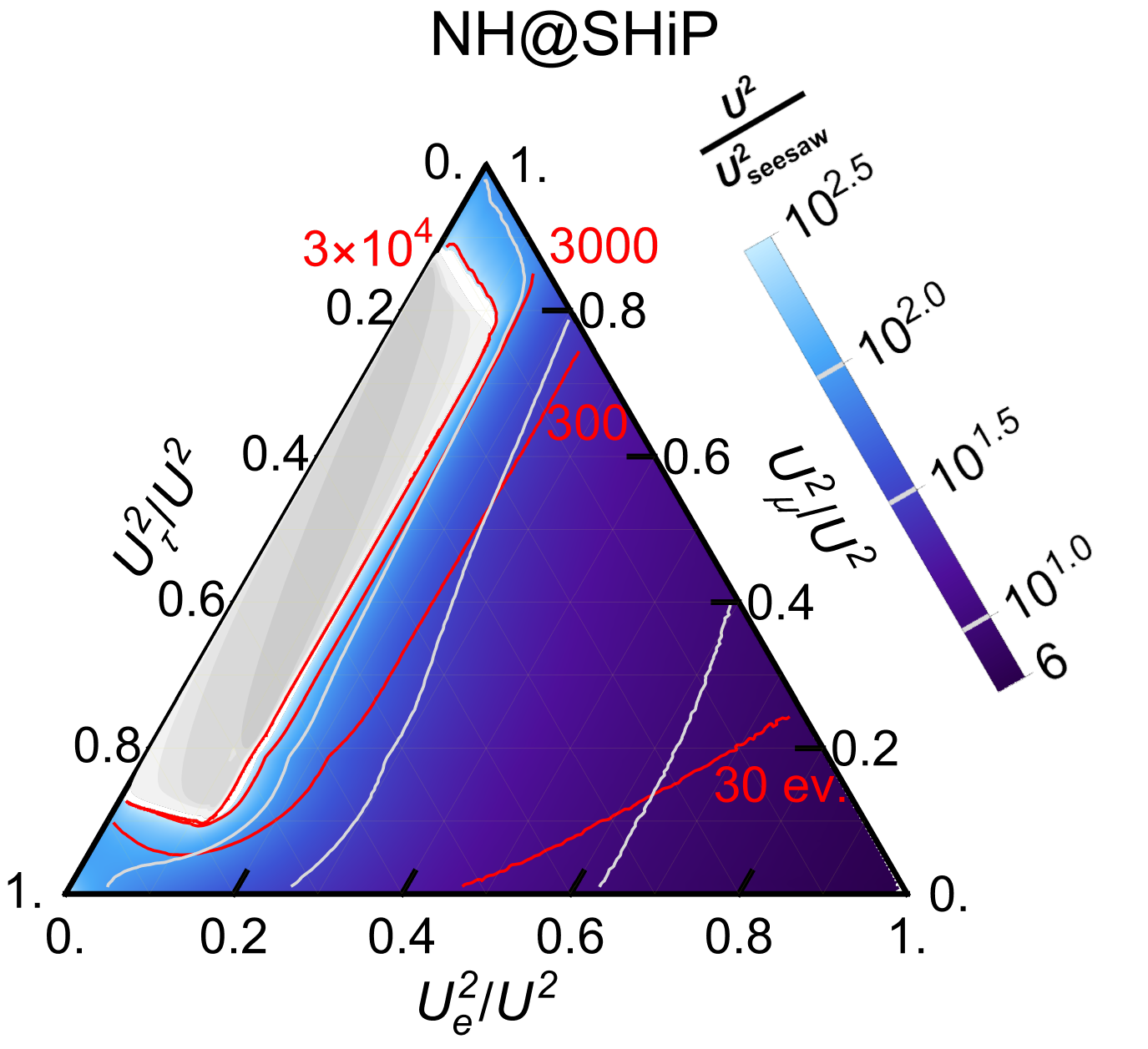}
    \includegraphics[width = 0.85\linewidth]{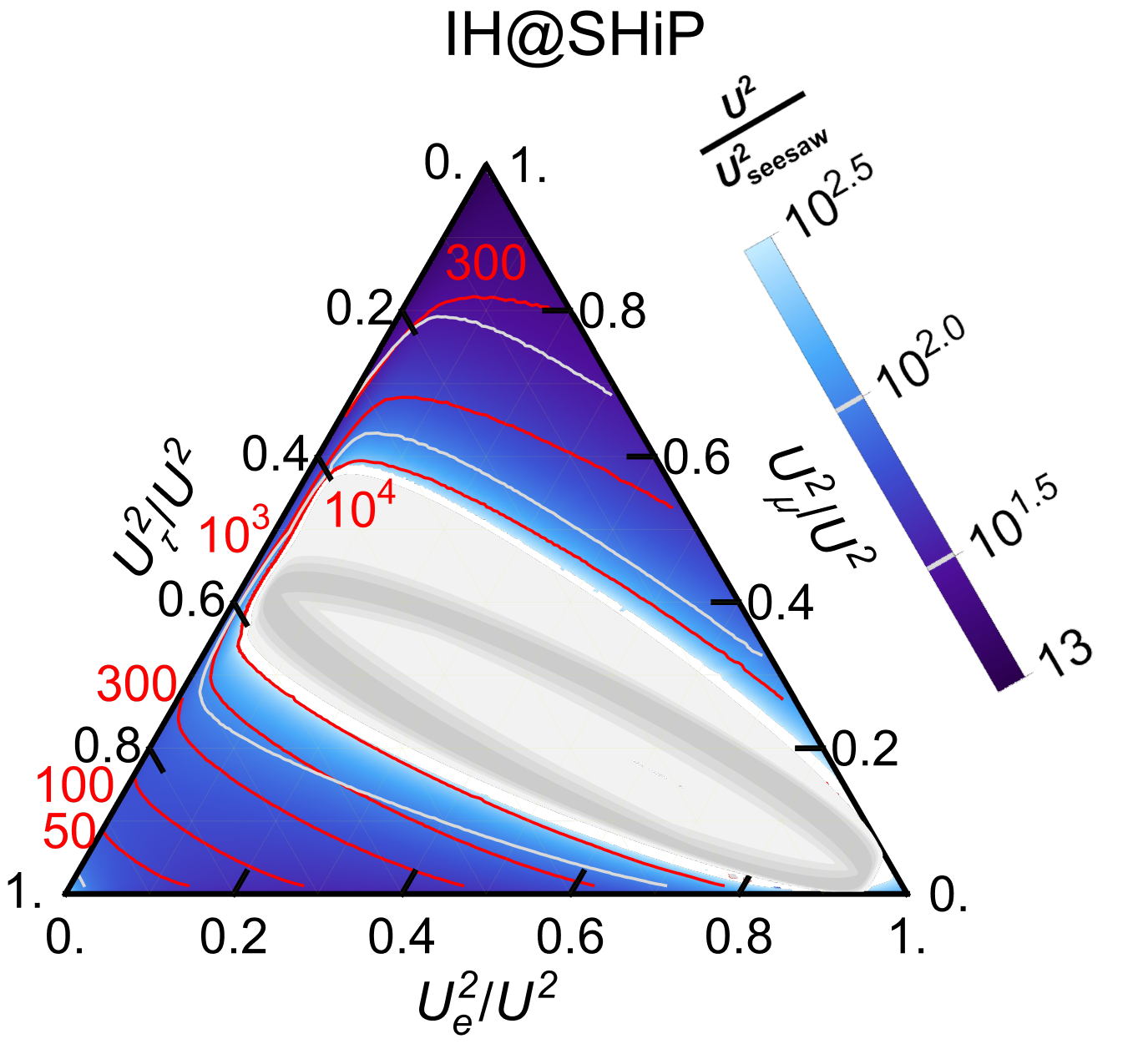}
    
    \caption{The exclusion reach of the SHiP experiment~\cite{SHiP_BDF_ECN3}, equivalent to the Fig.~\ref{fig:hierarchy_exclusion_maps}. Red lines are the corresponding isocontours of the number of \textbf{visible} events ($=N_\text{ev}\cdot \sum \text{br}_i$).}
    \label{fig:u2sens}
\end{figure}

The corresponding limits in terms of $U^2/U^2_\text{seesaw}$ for the SHiP experiment~\cite{SHiP_BDF_ECN3} are shown in Fig.~\ref{fig:u2sens}. For the specified mass, SHiP may be able to test the hypothesis of two seesaw HNLs within an order of magnitude above the seesaw bound.

\begin{figure}
    \centering
    \includegraphics[width = \linewidth]{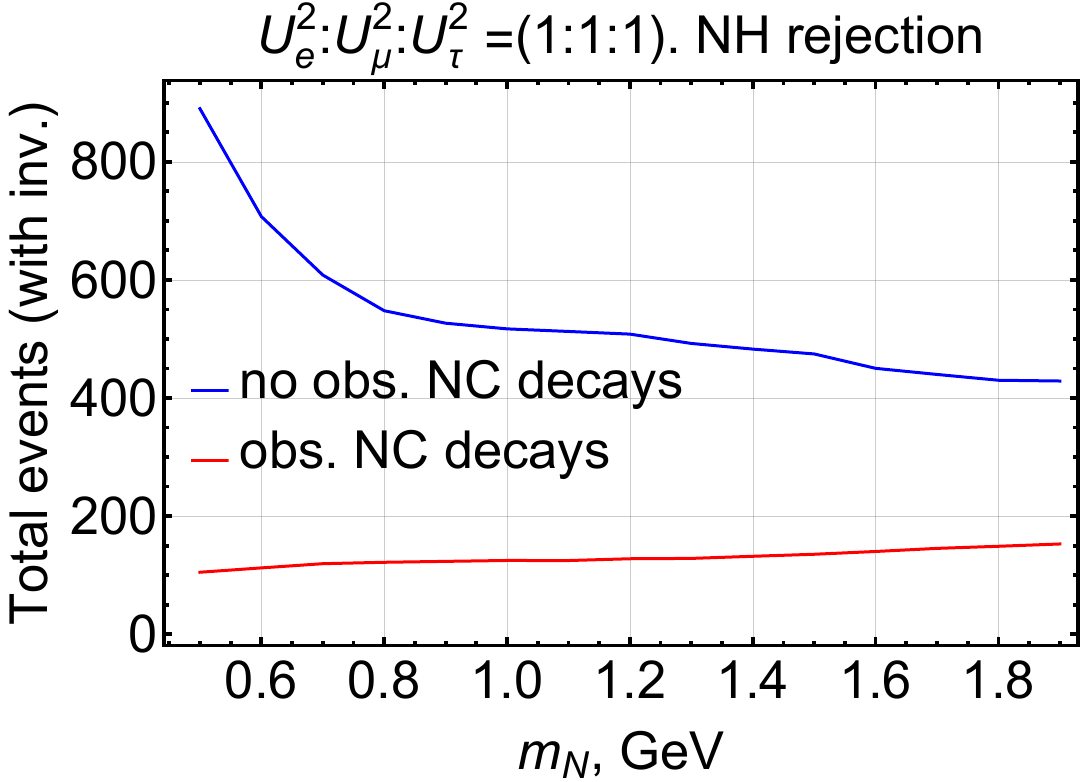}
    \caption{Number of events required to reject NH hypothesis for a model with equal mixing ratios $(x_e, x_\mu, x_\tau) = (1:1:1)$, consistent with IH, for two cases: while observing all visible decay modes (1-6) of Tab.~\ref{tab:decaymodes}, and while missing the NC mode (4)}
    \label{fig:NC_sensitivity}
\end{figure}

The discrimination power depends on the identification of decays into photons through the NC channel. Figure~\ref{fig:NC_sensitivity} illustrates how non-detection of the NC channel may increase the total number of decays by up to a factor of ten.

\section{Conclusions}
 
In this paper, we presented a method for differentiating between BSM models at the future Intensity Frontier experiments. Thanks to their ability to observe many events in the yet unexplored parameter space, they have the potential to infer the details of the underlying physics of the new particles.  

The main source of information relies on examining the decay modes of the particles. The portal models with a new scalar, fermion, vector, and photon-coupled axion-like particle may be distinguished with as few as up to ten events, owing to the distinct features of their decays. With more events, it becomes possible to further probe the details of the models. To list the possible implications, an abundant signal of new fermion particles allows to address its relation to the seesaw mechanism, shedding light on the paramount question of the origin of neutrino masses. The other examples include probing the model of an axial-like particle with general couplings to photons, gluons, and fermions, as well as exploring the non-minimality of a scalar/vector particle beyond the conventional portal model. In the general case, the new physics may be parametrized and probed in the framework of the effective field theory (EFT), without any theory-driven relations between the decay modes. This opens the possibility of exploring flavor physics and, for example, probe models with $B-L$ and $L_\mu-L_\tau$ coupled vector bosons~\cite{Langacker:2008yv}.

The procedure of ``probing the model'' refers to a measurement of its parameters under the assumption that the model itself is correct. The first necessary step is to identify the relevant classes of decay modes that are sensitive to the parameters. The relations between the branching ratios into some final states may be fixed by the structure of the model. These decay modes should be incorporated together into a specific class. When considered independently, they serve as a means to confirm the model's internal consistency and validate our starting presumption.

We employ our method to obtain the sensitivity of a generic experiment to the model of a pair of nearly degenerate sterile neutrinos that explain neutrino oscillations. The model is parametrized by the ratios of the mixing angles $x_\alpha = U^2_\alpha/U^2$, which are bound by neutrino oscillation data through the Casas-Ibarra parametrization, as depicted in Fig.~\ref{fig:neutrino_priors}. The relevant classes of decay modes are listed in Tab.~\ref{tab:decaymodes}.  Given an ample event count, a fermion particle with $x_\alpha$ inconsistent with the seesaw mechanism and a given hierarchy of neutrino mass may be identified as such. This opens up the possibility to differentiate between the two hierarchies, and perhaps even dismiss both, rendering the observed particle insufficient to explain the neutrino masses and invoking a further extension of the Standard Model. For common benchmark models of heavy neutral leptons, the necessary number of events is given in Tab.~\ref{tab:benchmark_exclusions}. Furthermore, with the same probe, it is possible to constrain the elusive Majorana phase of active neutrinos.

\newpage
\bibliographystyle{apsrev} 
\bibliography{bibliography}

\appendix
\begin{widetext}

\section{Software details}

The software is written in Python 3.10 and available as a module at \faGithub\href{https://github.com/omikulen/modeltesting}{ omikulen/modeltesting}. It contains utility functions for the procedure described in App.~\ref{sec:technical_appendix}, and a submodule \verb|modeltesting.neutrinos| for the analysis of sterile neutrinos.

\section{Neutrino priors}
\label{sec:neutrino_priors}

We consider an extension with two mass-degenerate HNLs in the approximate lepton symmetry limit $U^2\gg U^2_\text{seesaw} \equiv \sqrt{\Delta m^2_\text{atm}}/{m_N}$. In this case, the mixing angles of the two species are equal up to the small $|U^2_{\alpha 1} - U^2_{\alpha 2}|\sim U^2_\text{seesaw}$ correction. The relation between the mixing ratios and the neutrino oscillation parameters is given by the Casas-Ibarra parameterization~\cite{Casas:2001sr}:
\begin{equation}
    U^2_{\alpha} \propto m_i |V_{\alpha i}|^2 + m_j |V_{\alpha j}|^2 \pm 2\sqrt{m_i m_j} \,\text{Im}(V_{\alpha i} V_{\alpha j} e^{-i\eta} )
\end{equation}
with $(i, j) = (2, 3)$ and $(1,2)$ for NH and IH respectively. Here, $V$ is the PMNS matrix without the Majorana phase, which is included here explicitly as $\eta =  \eta_2$ (NH), $\eta_2 - \eta_1$ (IH) in the PDG convention~\cite{ParticleDataGroup:2020ssz}. The sign of the third term depends on the way of approximate lepton symmetry breaking. Since it cannot be constrained without explicitly observing the small deviations that violate the symmetry, each combination of $U^2_{\alpha}$ has two solutions for the Majorana phase: $\eta$ and $\eta+\pi$. The lightest neutrino mass ($m_1$ in NH and $m_3$ in IH) is set to zero.

NuFIT 5.2~\cite{Esteban:2020cvm} provides $\chi^2$ distributions for combinations of the neutrino parameters $\theta_{12}$, $\theta_{13}$, $\theta_{23}$, $\delta_\text{CP}$, $\Delta m^2_{12}$, $\Delta m^2_{3l}$. In this work, the results without the Super-Kamiokande data were used. For each pair of neutrino parameters $(p_1, p_2)$, we fix the remaining 4 parameters at their best-fit values, and compute the flavor ratios $(x_e, x_\mu)$ for the varying $p_1$, $p_2$, and the Majorana phase $\eta$.

\begin{table}[h!]
    \centering
    \begin{tabular}{|c|c|c|}
    \hline
         $p$ & $\quad$ NH range $\quad$ & $\quad$ IH range $\quad$
    \\
    \hline
         $\theta_{12}$, $^\circ$ & [30.7, 37.2] & [30.7, 37.2] \\ \hline
         $\theta_{13}$, $^\circ$ & [8.03, 9.06] & [8.07, 9.10] \\ \hline
         $\theta_{23}$, $^\circ$ & [38.6, 53.1] & [38.9, 53.1] \\ \hline
         $\Delta m^2_{12}$, $\unit[10^{-5}]{eV^2}$ & [6.55, 8.32] & [6.55, 8.32] \\ \hline
         $|\Delta m^2_{3l}|$, $\unit[10^{-3}]{eV^2}$ & [2.39, 2.64] & [2.37, 2.62] \\ \hline
         $\delta_\text{CP}$, $^\circ$ & [0, 360] & [150, 400]
    \\
    \hline
    \end{tabular}
    \caption{The regions in which the scan has been performed. For $p_1$, $p_2$, the scan step corresponds to 400 points in each range. For the Majorana phase $\eta$, the step corresponds to 5000 points in the $[0, 2\pi]$ interval.}
    \label{tab:neutrino_priors_details}
\end{table}

The resulting $(x_e, x_\mu)$ are put on a grid with $2.5 \cdot 10^{-3}$ step. Then, among all combinations of $(p_1, p_2, \eta)$ that lead to a specific point $(x_e, x_\mu)$, we choose the one with the smallest $\Delta \chi^2$ and associate this value with the point $(x_e, x_\mu)$. Finally, we translate $\Delta \chi^2$ into the prior probability as
\begin{equation}
    P = \exp(-\Delta\chi^2/2)
\end{equation}
corresponding to the assumption of $\chi^2$ following the 2-d.o.f. distribution.

\section{Technical details}

\label{sec:technical_appendix}

\subsection{Parametrization}

A physical model is parameterized by the total number $N$ of events that may happen within the acceptance of an experiment, and the branching ratios $\text{Br}_i$ for decays into $i$ pre-selected final states (channels). The branching ratios are known functions of the internal parameters of the model. The final states should be properly chosen to maximize the discriminating power in the space of models, as discussed in Sec.~\ref{sec:HNL}. 

The mean number of events in the $i$-th channel is given by
\begin{equation}
    \lambda_i = N\cdot \text{br}_i + b_i
\end{equation}
where $b_i$ is the background for this channel, independent on the model, and the effective branching ratio $\text{br}_i$ includes the average detection efficiency $\epsilon_i$ for the given set of decay products
\begin{equation}
    \text{br}_i\equiv \epsilon_i \text{Br}_i
\end{equation}

The details of the specific experiment are encoded in the efficiency coefficients $\epsilon_i$, background $b_i$, and the relation between the total number of events $N$ and the internal parameters (coupling constants) of the physical model.

\subsection{Hypothesis testing}

We define the procedure that determines whether the given experimental result $s_i$ contradicts with our hypothesis of some \textit{testable} model $t$ using the $\chi^2$ test. To test the consistency of the observed data with the model predictions $\lambda_i$, we do the following:
\begin{enumerate}
    \item Compute the quantity
\begin{equation}
    \chi_\text{exp.}^2(s_i | \lambda_i) = \sum_i \frac{(s_i - \lambda_i)^2}{\lambda_i}.
\end{equation}

    \item Build the cumulative probability distribution $\cdf(\chi^2_t)$ of the random variable $\chi_t^2 = \chi^2(s^t_i|\lambda_i)$ by drawing many realizations of random Poisson variables $s^t_i$ with mean values $\lambda_i$.
    \item Compute the $p$-value:
\begin{equation}
\label{eq:pval}
    p = \mathscr P_t \times \left(1 - \cdf(\chi^2_\text{exp.})\right),
\end{equation}
with $1 - \cdf(\chi^2_{\text{exp.}})$ being the probability that the random variable $\chi^{2}_t$ from the second step exceeds the experimental value $\chi^2_\text{exp.}$. This definition includes both the experimental confidence level and the prior probability of the physical model $\mathscr P_t \leq 1$.

    \item 
Finally, reject the model with confidence level CL, once the $p$-value is below the threshold $p< 1- \text{CL}$. 
\end{enumerate}

The approach of random realizations of the experimental results (instead of comparing with the analytic $\chi^2$-distribution) deals with two nuances. First, it allows one to properly define the threshold level for a Poisson variable with a small mean value $\sim 1$ when the approximation with the Gaussian distribution is not valid. Secondly, this approach handles a general case, when it is unknown beforehand how many of the specified decay channels actually contribute to $\chi^2$. 

\subsection{Sensitivity estimate}

For a real model, parameterized with branching ratios $\text{Br}_r$, the sensitivity for rejection a tested model with $\text{Br}_i$ is defined as the minimal number of events $N_r$ that results in the probability above the threshold $P$ to exclude the tested model at CL.

The input is:
\begin{itemize}
    \item $\text{Br}_r$, $\text{Br}_t$ for the real and tested model 
    \item experimental detection efficiency $\epsilon(=1)$ 
    \item background $b(=0)$
    \item prior probability of the tested model $\mathscr{P}_t (=1)$
\end{itemize}
where the numbers in the brackets represent the default values. The exact procedure employed in the software goes as follows:
\begin{enumerate}
    \item Make a starting rough estimate for the resulting $N_r$
    \begin{equation}
        N_r = \max\left[\frac{\lambda \chi_0}{\sum_i (\lambda \, \text{br}^t_i - \text{br}^r_i)^2/\text{br}^t_i}, \sqrt{\frac{\chi_0}{\sum_i (\text{br}^t_i - \text{br}^r_i/\lambda_i)^2/b_i}}\right], \qquad \lambda = \frac{\sum_i \text{br}^t_i}{\sum_i \text{br}^r_i}
    \end{equation}
    with two guesses in the brackets corresponding to background-free ($b_i \ll N_r \text{br}_i$) and background-dominated ($b_i \gg N_r \text{br}_i$) assumptions. The reference value $\chi^2_0$ is determined as 
    \begin{equation*}
        \cdf(\chi^2_0) = 1 - (1-\text{CL})/\mathscr P_t
    \end{equation*} for the analytical $\chi^2$ distribution with (number of channels - 1) degrees of freedom.
    \item For a given $N_r$, simulate $N_\text{samples} (= 10^4)$ realizations of experimental data $s_r$.
    \item For each set of $s_r$, find the best-fit number of events $N^\text{bf}_t$ for the tested model (as described in App.~\ref{sec:N-bf}) and compute the corresponding $\chi^2$. The data with $\chi^2$ exceeding some threshold $\chi^2_\text{th} (=30)$ is immediately deemed inconsistent with the tested model.
    \item  To further speed up the algorithm, we compute the average of the remaining $N^{\text{bf}}_t$ and build a single reference $\cdf$ for the model $(N^\text{bf}, \text{Br})_t$, instead of building a separate $\cdf$ for each dataset.
    \item Using the $p$-value from~\eqref{eq:pval}, find the fraction $\mathcal P_\text{rej.}$ of realizations that are inconsistent with the tested model $p<1-\text{CL}$.
    \item Using the gradient descent algorithm, adjust $N_r$:
    \begin{equation*}
        N_{r,\text{next}} = N_{r,\text{previous}} \times (1 + \alpha (\mathcal P_\text{rej.} - P) )  
    \end{equation*}
    with some convergence rate $\alpha(= 1)$. Return to step 2 and repeat until $\mathcal P_\text{rej.}$ converges to the desired probability threshold $P$, i.e. the error $|\mathcal P_\text{rej.} - \mathcal P|$ becomes small ($\leq 0.01$)
\end{enumerate}

\section{Finding $N^\text{bf}$}
\label{sec:N-bf}

In this appendix, we devise a numerical recipe to estimate $N^\text{bf}$ for a given dataset $s_i$ and a model with branching ratios $\text{br}_i$ and background $b_i$.

We formulate fitting as the minimization problem for the function
\begin{equation}
    \chi^2(N) = \sum_i \frac{[s_i - (b_i + N \text{br}_i) ]^2}{b_i + N \text{br}_i}
\end{equation}
which yields the following polynomial equation of degree $2\cdot i$
\begin{equation}
    \frac{d\chi^2}{dN} = \sum_i \text{br}_i \left[ \frac{s^2}{(b_i + N \text{br}_i)^2} - 1\right] = 0.
\end{equation}
Solving it numerically may be time-consuming; moreover, the algorithm should be stable to work consistently for arbitrary input $s_i$, $b_i$, $\text{br}_i$. To avoid these difficulties, we make the following simplification: let us denote $\Delta s_i = s_i - b_i$
\begin{equation}
    \chi^2(N) = \sum_i \frac{[\Delta s_i - N \text{br}_i]^2}{b_i + N \text{br}_i}
\end{equation}
We consider an approximate $\chi^2$, where we split the sum into two:
\begin{align}
    &\chi^2_\text{a} = \sum_1 \frac{(\Delta s_i - N \text{br}_i)^2}{b_i} + \sum_2 \frac{(\Delta s_i - N \text{br}_i)^2}{N \text{br}_i} 
    \\
    &\sum_1 = \sum_{\Delta s_i < b_i}, \qquad \sum_2 = \sum_{\Delta s_i > b_i}
\end{align}
where we assume that $\Delta s_i \sim N\text{br}_i$, and only the dominant of the two terms in the denominator is left in each of the two sums. This yields the following equation:
\begin{equation}
    C_1 N^3 + (C_2 - C_3) N^2 - C_4 = 0 
\end{equation}
\begin{align*}
    C_1 = 2 \sum_1 \frac{\text{br}_i^2}{b_i} \qquad
    C_2 = \sum_2 \text{br}_i \qquad  C_3 =  2 \sum_1 \frac{\Delta s_i \text{br}_i}{b_i} \qquad 
    C_4 = \sum_2 \frac{\Delta s_i^2}{\text{br}_i}
\end{align*}
In the two limiting cases, when all $\Delta s_i > b_i$ or all $\Delta s_i < b_i$ such that only one sum survives, the answer simplifies to:
\begin{align*}
    &N = \sqrt{\frac{C_4}{C_2}} = \sqrt{\frac{\sum \Delta s_i^2/\text{br}_i}{\sum \text{br}_i}}, \qquad \forall \Delta s_i > b_i \\
    &N = \frac{C_3}{C_1} = 
\frac{\sum \Delta s_i \text{br}_i / b_i}{\sum \text{br}^2_i / b_i}, \qquad \,\,\forall\Delta s_i < b_i
\end{align*}

In the general case, the solution is given by
\begin{equation*}
    N = \left(\frac{(\sqrt{B} + \sqrt{B - 4 A^3})^2}{4}\right)^{\frac{1}{3}} + A^2 \left(\frac{(\sqrt{B} + \sqrt{B - 4 A^3})^2}{4}\right)^{-\frac{1}{3}} - A, \qquad B\geq  4 A^3
\end{equation*}
\begin{equation*}
    N = A \left[ 2 \cos \left(\frac{2}{3} \arctan \frac{\sqrt{4 A^3 - B}}{\sqrt{B}}\right) - 1\right], \qquad B < 4A^3
\end{equation*}
which are the real positive roots of the equation $N^3 + 3A N^2 - B = 0$, with $B = \frac{C_4}{C_1}$ and $A = \frac{C_2-C_3}{3C_1}$

\end{widetext}

\end{document}